\documentclass[12pt, a4paper]{article}
\pdfoutput=1
\usepackage{amsmath}
\usepackage{amssymb}
\usepackage{graphicx}
\usepackage{subcaption}
\usepackage{color}
\usepackage{url, cite}
\usepackage[sort&compress, numbers, merge]{natbib}

\definecolor{purple}{rgb}{0.5 ,0, 0.7}
\definecolor{bluegreen}{rgb}{0, 0.45, 0.35}
\usepackage{hyperref}
\hypersetup{colorlinks=true, linkcolor=bluegreen, citecolor=purple, urlcolor=blue}

\allowdisplaybreaks

\begin{document}

\begin{titlepage}
\begin{center}
\leavevmode \\
\vspace{ 0cm}

\hfill UT-16-10 \\
\hfill APCTP Pre2016-005

\noindent
\vskip 2 cm
{\Huge  Generalized Pole Inflation:} \\
\vskip 4 mm
 {\Large  Hilltop, Natural, and Chaotic Inflationary Attractors}

\vglue .6in

{\large
 Takahiro Terada
}

\vglue.3in

\textit{
 Department of Physics, The University of Tokyo, Tokyo 113-0033, Japan,  and \\
   Asia Pacific Center for Theoretical Physics, Pohang 37673, South Korea\\
}

\end{center}

\vglue 0.7in

\begin{abstract}
A reformulation of inflationary model analyses appeared recently, in which inflationary observables are determined by the structure of a pole in the inflaton kinetic term rather than the shape of the inflaton potential.
We comprehensively study this framework with an arbitrary order of the pole taking into account possible additional poles in the kinetic term or in the potential.
Depending on the setup, the canonical potential becomes the form of hilltop or plateau models, variants of natural inflation, power-law inflation, or monomial/polynomial chaotic inflation.
We demonstrate attractor behaviors of these models and compute corrections from the additional poles to the inflationary observables.
\end{abstract}
\end{titlepage}

%%%%%%%%%%%%%%%%%%%%%%%%%%%%%%%%%%%%%%%%%

\section{Introduction}
%inflation
Due to the recent detection of gravitational waves~\cite{Abbott:2016blz}, validity of General Relativity has been extended to a new frontier~\cite{TheLIGOScientific:2016src}.  On the other hand, evidence of gravitational waves from the inflationary period~\cite{Starobinsky:1980te, Sato:1980yn, Guth:1980zm, Linde:1981mu, Albrecht:1982wi, Linde:1983gd}, which was invented to solve the homogeneity, flatness, and monopole puzzles, has yet to be observed.  
In addition to the gravitational fluctuation, scalar perturbation is generated during inflation as quantum fluctuation of inflaton field, and it is transferred to curvature perturbation eventually leading to the large scale structure of our universe.
Its almost scale-invariant, adiabatic, and Gaussian features have been well established by cosmic microwave background (CMB) observations~\cite{Bennett:2012zja, Hinshaw:2012aka, Ade:2015xua, Ade:2015lrj}.
The deviation from scale invariance is parametrized by the scalar spectral index $n_{\text{s}}=0.9666\pm 0.0062$ (68\% CL, Planck TT+lowP; $\Lambda$CDM model with tensor)~\cite{Ade:2015xua, Ade:2015lrj}, and the upper bound on the tensor-to-scalar ratio becomes tighter and tighter, $r < 0.07$ (95\% CL; Planck, BICEP2/Keck-Array combined)~\cite{Array:2015xqh} at the pivot scale $k=0.05\, \text{Mpc}^{-1}$.

%pole inflation
Meanwhile, there are many inflationary models predicting the observable values at the sweet spot of the Planck constraint contour: the Starobinsky model~\cite{Starobinsky:1980te} and Higgs inflation model with non-minimal coupling to gravity~\cite{Bezrukov:2007ep} are two notable examples, whose predictions are $n_{\text{s}}= 1 - 2/N$ and $r=12/N^2$ where $N$ is the $e$-folding number.
In these models, the approximate shift symmetry in terms of the canonical inflaton in the Einstein frame is originated from scale invariance in the Jordan frame action.
Generalization of Higgs inflation is further studied under the name of universal attractor (also called $\xi$-attractor or induced inflation)~\cite{Kallosh:2013maa, Kallosh:2013tua, Giudice:2014toa, Pallis:2014dma, Kallosh:2014laa, Mosk:2014cba, Pieroni:2015cma}.
The name ``attractor'' refers to the fact that this class of models predicts the same values for $n_{\text{s}}$ and $r$ at the limit of strong coupling, $\xi \gg 1$, irrespectively of an arbitrary function characterizing the model.
There is another branch of attractors called $\alpha$-attractor~\cite{Ellis:2013nxa, Ferrara:2013rsa, Kallosh:2013yoa, Kallosh:2014rga, Carrasco:2015pla, Roest:2015qya, Linde:2015uga, Scalisi:2015qga}, which generalizes conformal attractor~\cite{Kallosh:2013hoa}, intimately related to the geometric properties of supergravity with K\"{a}hler manifold whose curvature is inversely proportional to the parameter $\alpha$ \footnote{In a closely related formulation~\cite{Ozkan:2015kma}, $(\alpha-1)$ can be interpreted as a parameter measuring how much conformal symmetry is broken to scale symmetry in the underlying theory.} \cite{Kallosh:2015zsa, Carrasco:2015uma}.
This also predicts the same value for $n_{\text{s}}$, and $r$ is given by $r=12\alpha / N^2$, in the small $\alpha$ limit and independently of the details of the potential.
In some cases, these models have another attractor point in the opposite limit $\alpha \to \infty$ or $\xi \to 0$ where the prediction coincides with that of chaotic inflation~\cite{Linde:1983gd} with a quadratic potential~\cite{Kallosh:2014rga, Kallosh:2014laa, Mosk:2014cba}.
This simply reflects the fact that expansion of generic potentials at the minimum starts from the quadratic term.  This mechanism is called double attractor.

These attractors can be understood in a unified manner noticing the fact that their actions in the Einstein frame are characterized by a second order pole in the coefficient of the inflaton kinetic term like $- a_p / (2( \varphi-\varphi_0)^p) (\partial_{\mu} \varphi )^2$ with $p=2$ where $\varphi_0$ is the location of the pole~\cite{Galante:2014ifa}.
In fact, $\alpha$-attractor and a part of $\xi$-attractor are equivalent with the identification $\alpha=1+1/(6\xi)=2 a_2 /3$~\cite{Galante:2014ifa}.
Upon canonical normalization, the inflaton field is exponentially stretched out, and the potential becomes exponentially flat.
This can be viewed as an extreme case of running kinetic inflation~\cite{Takahashi:2010ky, Nakayama:2010kt}.
Remarkably, in the limit of small $\alpha$, the spectral index is determined solely by the order of the pole while the tensor-to-scalar ratio is also controlled by the residue of the pole~\cite{Galante:2014ifa}.
Cases of higher order poles ($p\geq 2$) and its relation to shift symmetry and its soft breaking was discussed in Ref.~\cite{Broy:2015qna}, where this paradigm was called pole inflation.
These pole inflation models with $p\geq2$ have a plateau-type potential which asymptotes to a constant, and predict $1-2/N \leq n_{\text{s}}<1-1/N$ in the attractor limit $a_p \to 0$.
The other cases ($p<2$) were briefly mentioned in Refs.~\cite{Galante:2014ifa, Broy:2015qna}.

Inflationary attractors beyond the Einstein gravity are also studied in the literature.
A duality relation between superconformal $\alpha$-attractor and higher curvature supergravity was elaborated in Ref.~\cite{Cecotti:2014ipa}.
An $f(R)$ gravity generalization of the Starobinsky model as another inflationary attractor was discussed in Ref.~\cite{Rinaldi:2015yoa, *Rinaldi:2015ana}.
In the Einstein frame, its form is of pole inflation with $p=2+\epsilon$ $(|\epsilon | \ll 1)$, but its scalar potential also has a pole of order $(p-2)/2$ and a logarithmic singularity.
Its prediction of $n_{\text{s}}$ and $r$ is close to that of chaotic inflation model with a linear potential.\footnote{
The attraction to the linear potential in the strong coupling limit may be a common property of generic theories with scale invariance broken by loop corrections (logarithmic functions).  See \textit{e.g.}~Refs.~\cite{Rinaldi:2015yoa, Kannike:2015kda}.
}
More general orders of the pole in the potential were mentioned briefly.

%this paper
In this paper, we extend these studies of pole inflation and obtain various inflationary potentials.
First, we review pole inflation in Sec.~\ref{sec:fractional} and demonstrate attractor behaviors of pole inflation with various pole orders $p$ for the first time for $p\neq 2$.
This includes the so-called hilltop model~\cite{Linde:1981mu, Albrecht:1982wi, Boubekeur:2005zm} in the case of $p<2$.
In Sec.~\ref{sec:first}, we consider the case of first order pole inflation ($p=1$). 
As we will see, in a concrete setup, the first order pole inflation includes variants of the natural inflation model~\cite{Freese:1990rb, Adams:1992bn}.
We comment on the validity of the effective field theories motivating the study of additional poles either in the kinetic term or in the potential.
We consider corrections from such additional poles to the inflationary observables in Sec.~\ref{sec:perturbation}.
This generalizes the discussion on shift symmetry breaking in Ref.~\cite{Broy:2015qna}. 
In Sec.~\ref{sec:singular}, we consider presence of unsuppressed poles of arbitrary orders both in the kinetic term and the potential, and obtain monomial potentials for chaotic inflation.
We show double attractor behavior of some examples and work out corrections to the inflationary observables for this case as well, generalizing the findings in Refs.~\cite{Broy:2015qna, Rinaldi:2015yoa}.
These new inflationary attractor models are qualitatively different from the conventional attractors and pole inflation. 
In this paper, the reduced Planck unit is taken, $c= \hbar = M_{\text{P}}/\sqrt{8\pi}=1$.

%%%%%%%%%%%%%%%%%%%%%%%%%%%%%%%%%%%%%%%%%
\section{Pole inflation and its attractor behavior \label{sec:fractional}}
%review and formulae
First, we review pole inflation~\cite{Galante:2014ifa, Broy:2015qna} and list relevant formulae.
Let us begin with a Jordan frame Lagrangian with a non-canonical scalar field $\widetilde{\varphi}$,
\begin{align}
\mathcal{L}=\sqrt{-g_{\text{J}}}\left[ \frac{1}{2}\Omega_{\text{J}}(\widetilde{\varphi})R_{\text{J}}-\frac{1}{2}K_{\text{J}}(\widetilde{\varphi}) g^{\mu \nu}_{\text{J}} \partial_{\mu}\widetilde{\varphi}\partial_{\nu}\widetilde{\varphi} -V_{\text{J}}(\widetilde{\varphi}) \right],
\end{align}
where the subscript J denotes the Jordan frame variables.
Applying the Weyl transformation, $g_{\mu \nu}^{\text{J}}=\Omega_{\text{J}}^{-1}g_{\mu \nu}^{\text{E}}$, 
it is expressed in terms of the Einstein frame variables up to a surface term,
\begin{align}
\mathcal{L}=\sqrt{-g_{\text{E}}}\left[ \frac{1}{2}\Omega_{\text{E}}(\widetilde{\varphi})R_{\text{E}}-\frac{1}{2}K_{\text{E}}(\widetilde{\varphi}) g^{\mu \nu}_{\text{E}} \partial_{\mu}\widetilde{\varphi}\partial_{\nu}\widetilde{\varphi} -V_{\text{E}}(\widetilde{\varphi}) \right],
\end{align}
where $\Omega_{\text{E}}(\widetilde{\varphi})\equiv 1$, and 
\begin{align}
K_{\text{E}}(\widetilde{\varphi}) =& \frac{K_{\text{J}}(\widetilde{\varphi})}{\Omega_{\text{J}}(\widetilde{\varphi})}+ \frac{3 \Omega'{}^{2}_{\text{J}}(\widetilde{\varphi})}{2 \Omega^2_{\text{J}}(\widetilde{\varphi})},  & \text{and} & &  V_{\text{E}}(\widetilde{\varphi})=&\frac{V_{\text{J}}(\widetilde{\varphi})}{\Omega^2_{\text{J}}(\widetilde{\varphi})}. \label{frames_relation}
\end{align}
In the following, we focus on the Einstein frame and omit the gravity part and the subscript E in the Lagrangian.

Suppose there is a point in the inflaton field space where the kinetic term becomes singular.
Redefine the origin of the field, $\widetilde{\varphi}\to \varphi=\varphi (\widetilde{\varphi})$, in such a way that the singular point coincides with the origin of the field.
At the point, we expand the kinetic term as a Laurent series.
Expressing the order of the highest relevant pole as $p$, the Lagrangian is given by
\begin{align}
\left( \sqrt{-g}\right)^{-1}\mathcal{L}= - \frac{a_p}{2 \varphi^p}   \partial^{\mu} \varphi \partial_{\mu} \varphi - V_0 \left( 1 -c \varphi +\mathcal{O}(\varphi^2)  \right), \label{Lpole}
\end{align}
where the potential is assumed to be regular at the origin and expanded as a Taylor series.
$V_0$ is an overall coefficient which can be used to fit the amplitude of the curvature perturbation.
The field sign and normalization can be chosen in such a way that $c=+1$ in the potential, which implies inflation occurs in the side of $\varphi>0$.
We can recover the general case by replacing $a_{p}$ with $a_{p}c^{p-2}$ in the following expressions.
We assume $a_p>0$ to avoid the negative-norm state.
Even if lower order poles coexist at the origin, their effects are subdominant near the origin, where inflation is supposed to occur, unless their coefficients are too large.

Around the pole, the canonical inflaton $\phi$ is obtained as
\begin{align}
\phi = \begin{cases}
  \frac{2\sqrt{a_p}}{p-2}\varphi^{-\frac{p-2}{2}}  & (p\neq 2), \\
 -\sqrt{a_p} \log \varphi  & (p=2),
 \end{cases}
\end{align}
 up to an integration constant. 
 The canonical inflaton potential is therefore
 \begin{align}
 V= \begin{cases}
 V_0 \left( 1 - \left( \frac{p-2}{2\sqrt{a_p}}\phi \right)^{-\frac{2}{p-2}} +\cdots \right)  & (p \neq 2), \\
 V_0 \left(1 - e^{-\phi/\sqrt{a_p}} +\cdots \right) & (p=2), 
 \end{cases}  \label{Vcan}
 \end{align}
 where dots represent subdominant terms in the large field region in terms of $\phi$, which corresponds to the region near the pole of $\varphi$.
 The $e$-folding number in terms of the field value is
 \begin{align}
 N=
 &\frac{a_p}{p-1} \left( \frac{1}{\varphi_N^{p-1}} - \frac{1}{\varphi_{\text{end}}^{p-1}} \right) \qquad (p \neq 1), \label{Ne}
 \end{align}
where $\varphi_N$ is the field value corresponding to $N$ $e$-foldings and $\varphi_{\text{end}}=(2a_p)^{\frac{1}{p}}$ is the field value when inflation ends, \textit{i.e.}  the slow-roll parameter $\epsilon$ becomes one.
The spectral index and tensor-to-scalar ratio are calculated as~\cite{Galante:2014ifa}
\begin{align}
n_{\text{s}}=&1-\frac{p}{(p-1)N}, & r = & \frac{8}{a_p} \left( \frac{a_p}{(p-1)N} \right)^{\frac{p}{p-1}},  \label{nsr_p}
\end{align}
for $p \neq 1$ at the lowest order in $N^{-1}$.
Note that $n_{\text{s}}$ is independent of $a_p$ at this order, and it is determined solely by the order of the pole $p$.  On the other hand, $r$ depends also on $a_p$, but its dependence becomes week in the large $p$ limit.

\begin{figure}[htb]
 \centering
  \subcaptionbox{original potential.\label{sfig:sc_pt_org}}
{\includegraphics[width=0.32\columnwidth]{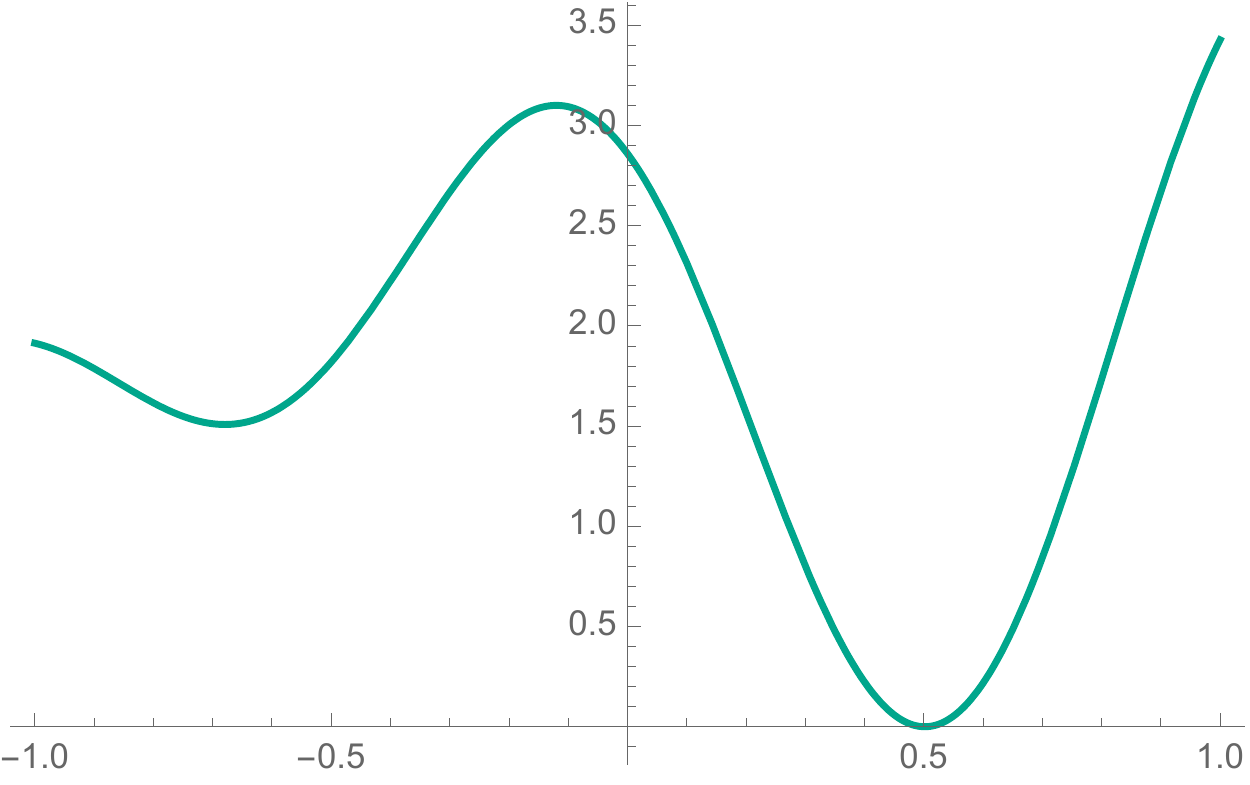}}~
 \subcaptionbox{$p<2$: hilltop.\label{sfig:sc_pt_hill}}
{\includegraphics[width=0.32\columnwidth]{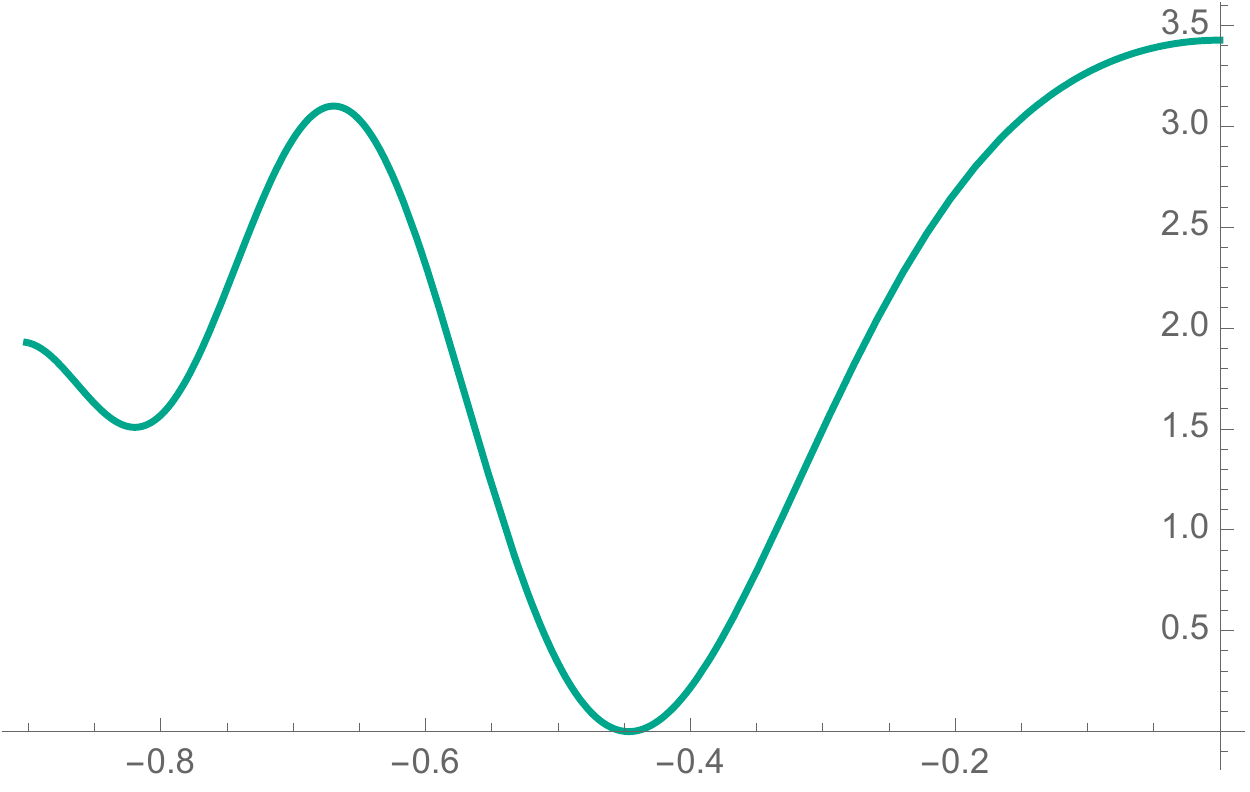}}~
 \subcaptionbox{$p\geq 2$: inverse-hilltop. \label{sfig:sc_pt_invhill}}
{\includegraphics[width=0.32\columnwidth]{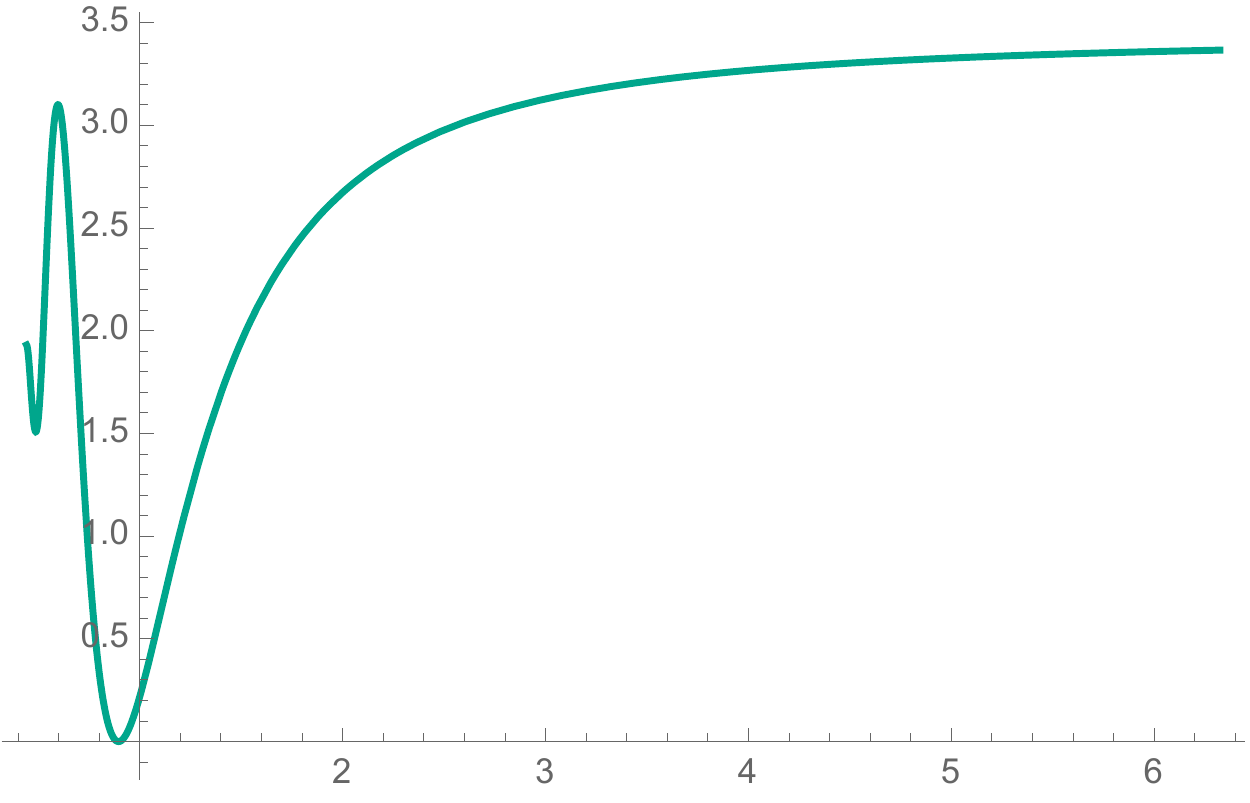}}~

  \caption{Effects of canonical normalization.
  (\subref{sfig:sc_pt_org}) The original potential before canonical normalization.
  (\subref{sfig:sc_pt_hill}) The canonical potential for $p<2$, which has a finite flat region and realizes hilltop inflation.  (\subref{sfig:sc_pt_invhill}) The canonical potential for $p\geq 2$, which is Starobinsky-like ($p=2$) or inverse-hilltop ($p>2$).}
 \label{fig:sc_pt}
  \end{figure}

We can in principle extend the definition of $p$ into non-integer values, and consider the cases with $p<2$ as well as $p\geq 2$.
For $p\geq 2$, the place of the pole $\varphi=0$ corresponds to $\phi\to \infty$ in terms of the canonical field, 
but this becomes $\phi=0$ for $0<p<2$.
In the latter case, the inflaton rolls down on the hill to the negative side, $\phi<0$, and the hilltop inflation occurs there.
This case is also an attractor in the sense that generic potentials are deformed into the hilltop shape in the limit $a_p \to 0$, and the predictions are attracted to eq.~\eqref{nsr_p}.
These facts are visualized schematically in Fig.~\ref{fig:sc_pt}.
The field excursion during inflation is estimated as (\textit{cf.}~Ref.~\cite{Garcia-Bellido:2014wfa})
\begin{align}
\Delta \phi \simeq \begin{cases}
\phi_N = \frac{2}{p-2}a_p^{\frac{1}{2(p-1)}}\left( (p-1) N \right)^{\frac{p-2}{2(p-1)}}   &  (p>2) , \\
\sqrt{a_p} \log\left( 1 +  \sqrt{\frac{2}{a_p}}N \right) & (p=2), \\
|\phi_{\text{end}}|= \frac{1}{\sqrt{2}(2-p)}(2 a_p )^{\frac{1}{p}}   &  (0<p<2).
\end{cases}
\end{align}
Note that the dependence on $a_p$ is strongest in the last case, and it does not depend on a positive power of  $e$-folding number.  Thus, compared to the other cases, pole inflation with $0<p<2$ tends to be (but not necessarily) small-field inflation.
For $0<p<1$, eq.~\eqref{Ne} becomes $N=\frac{a_p}{1-p} \left( \varphi_{\text{end}}^{1-p} - \varphi_N^{1-p}\right) \simeq \frac{a_p}{1-p} \varphi_{\text{end}}^{1-p}$ implying $n_{\text{s}}$ and $r$ depends weakly on $\varphi_N$.
Taking $\varphi_N \to 0$, we have $n_{\text{s}} \to - \infty$ and $r \to 0$, and we no more consider this case.
The case $p=1$ is separately discussed in Sec.~\ref{sec:first}.

With the above formulae, we can constrain the value of $p$ in the attractor limit $a_p \to 0$.
Precise values of constraints depend sensitively on the used data sets and assumptions.
Using the one mentioned at the Introduction, $n_{\text{s}}=0.9666\pm 0.0062$, the constraints turn out to be\footnote{The right hand side of the lower left constraint is weak because the observational constraint on $n_{\text{s}}$ is close to 0.98, while the predictions of pole inflation for $N=50$ asymptotes to 0.98 in the limit $p\to \infty$.}
\begin{align*}
2.02<p< 3.78  &  \quad (68\% \text{ CL}, N=50) , &  1.73<p<2.58 & \quad (68\% \text{ CL}, N=60), \\
1.78<p< 21.0  & \quad  (95\% \text{ CL}, N=50) , & 1.57<p<4.85  & \quad  (95\% \text{ CL}, N=60).
\end{align*}
In terms of the power $n= -2/(p-2)$ in the canonical potential, $V-V_0 \propto \phi^{n}$ (see eq.~\eqref{Vcan}), these constraints read
\begin{align*}
-98.0<n<-1.13 &  \quad (68\% \text{ CL}, N=50) , &  n>7.32, \, n<-3.43 & \quad (68\% \text{ CL}, N=60), \\
n>8.90, \, n<-0.105  & \quad  (95\% \text{ CL}, N=50) , & n>4.67, \, n<-0.703  & \quad  (95\% \text{ CL}, N=60).
\end{align*}

%demonstration with an example
Let us demonstrate the attractor behavior of pole inflation taking a monomial potential as a simplest yet illustrative example.
Inspired by the kinetic term of the superconformal $\alpha$-attractor~\cite{Kallosh:2013yoa},
\begin{align}
-K_{\Phi \bar{\Phi}} \partial^{\mu}\bar{\Phi}\partial_{\mu}\Phi = - \frac{3 \alpha}{ \left( 1 - |\Phi|^2\right)^2} \partial^{\mu}\bar{\Phi}\partial_{\mu}\Phi,
\end{align}
which is derived from the K\"{a}hler potential $K= - 3\alpha \log ( 1 - |\Phi|^2)$, 
consider the following model,
\begin{align}
\left( \sqrt{-g}\right)^{-1}\mathcal{L}= - \frac{a_p}{2(1 - \widetilde{\varphi}^2)^p} \partial^{\mu} \widetilde{\varphi} \partial_{\mu} \widetilde{\varphi} - \lambda_m \widetilde{\varphi}^m. \label{L_p}
\end{align}
%We introduced a new variable $\widetilde{\varphi}$ to distinguish it from $\varphi$ in eq.~\eqref{Lpole}, which has a pole at the origin, and the canonical field $\phi$. 
It has first order poles at $\widetilde{\varphi}=\pm 1$.
The potential has been taken as a monomial for simplicity.
Taken at its face value, the potential becomes negative for odd $m$, but it should be assumed that it represents the approximate form of the potential where inflation happens (near $\widetilde{\varphi} = 1$).
It is implicitly assumed that the potential near the origin $(\widetilde{\varphi} < \widetilde{\varphi}_{\text{end}})$ allows a stable minimum with an almost vanishing positive cosmological constant.

\begin{figure}[htbp!]
\centering
\includegraphics[width=0.75\columnwidth]{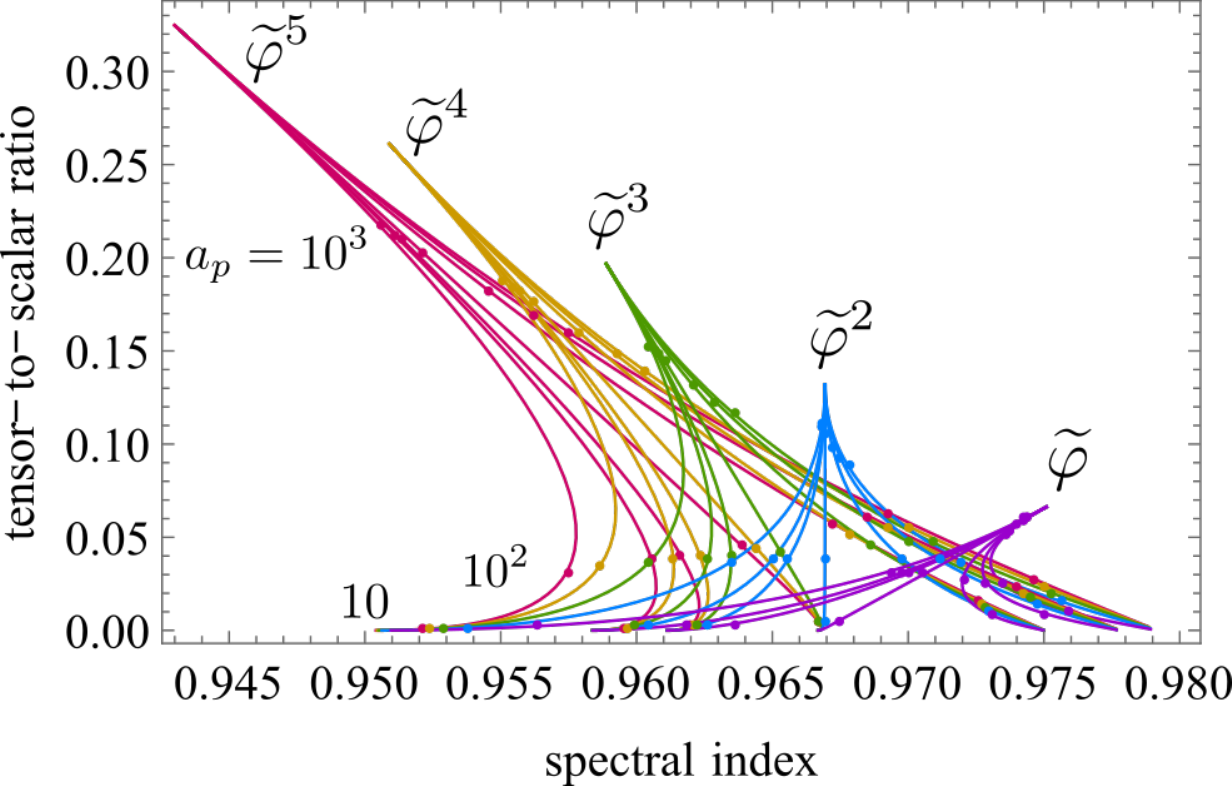}
\vskip 3 mm
\includegraphics[width=0.75\columnwidth]{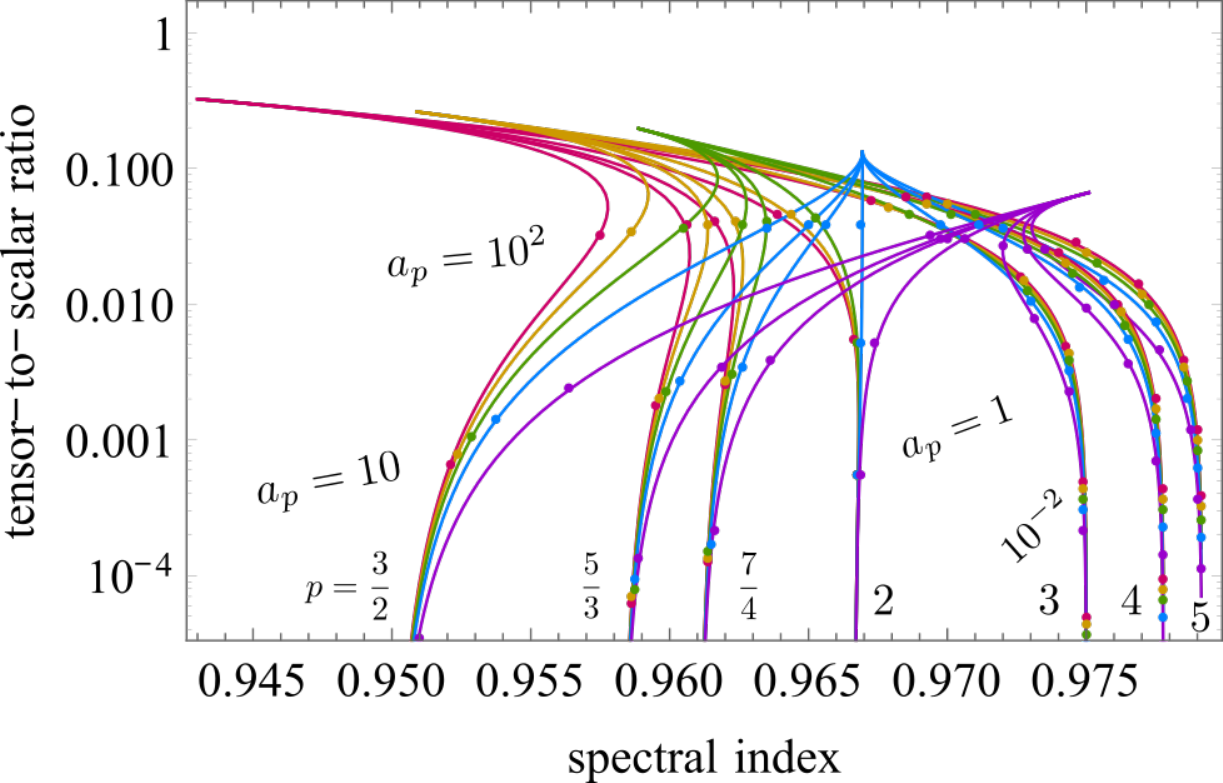}
  \caption{Attractor behavior of pole inflation in the linear and logarithmic scales.
  At the top of the Figures, lines start at the points of the predictions of the original monomial potentials~\eqref{L_p} with power $m=1, \, 2, \, 3, \, 4$, and 5 from right to left.
    At the bottom of the Figures, the horizontal values of these lines are attracted to the attractor values~\eqref{nsr_p} for $p= 3/2, \, 5/3, \,  7/4, \, 2, \, 3, \, 4,$ and 5 from left to right.
    Dots correspond to $a_p= 10^3, \, 10^2,$ and $10$ from top to bottom in the top Figure and $a_p = 10^2, \, 10, \, 1, \, 10^{-2}, \, 10^{-4},$ and $10^{-6}$ from top to bottom in the bottom Figure.
    The $e$-folding is set to $N=60$.
    For the most right lines, we show up to $a_5= 10^{-7}$.  It goes below further in the limit $a_5 \to 0$.
  }
 \label{fig:fractional}
  \end{figure}

The canonical inflaton is given by a hypergeometric function, $\phi=\sqrt{a_p} \widetilde{\varphi} \, {}_{2}F_{1}\left(\frac{1}{2},\frac{p}{2}; \frac{3}{2}; \widetilde{\varphi}^2 \right)$.
We cannot explicitly invert this function for general $p$, but inflationary observables can be calculated in the basis of the original variable.
The attractor behavior of this model for various $p$ and $m$ are shown in Fig.~\ref{fig:fractional}.
Poles of different order constitute a series of inflationary attractors.
Poles with $p>2$ predict higher values of $n_{\text{s}}$ than that of $\alpha$-attractor, and the maximum value is $n_{\text{s}}=1-1/N$ for $p\to \infty$.
Poles with $p<2$ predict lower values of $n_{\text{s}}$.
Interestingly, curves in Fig.~\ref{fig:fractional} become narrow in the horizontal direction at an intermediate value of $a_p$ where $r$ is about $10^{-2}$ to $10^{-1}$.
If this is to explain why the observational value of $n_{\text{s}}$ is what we see, the tensor mode will be detected in the near future by observation of CMB $B$-mode polarization.
If the tensor mode is not found and constraints on $n_{\text{s}}$ become tighter, it will help us identify the order of the pole $p$.

%%%%%%%%%%%%%%%%%%%%%%%%%%%%%%%%%%%%%%%%%
\section{First order pole inflation \label{sec:first}}

We turn to the case of first order pole inflation, \textit{i.e.}~$p=1$.
In this case, the canonical potential~\eqref{Vcan} is $V=V_0 ( 1 - \phi^2 /(4a_1 ) +\cdots )$, and eq.~\eqref{Ne} is replaced with
\begin{align}
N =& a_p \log \left( \frac{\varphi_{\text{end}}}{\varphi_N} \right) \qquad (p=1). \label{Ne1}
\end{align}
In this expression, one cannot neglect the contribution of $\varphi_{\text{end}}$.
If the potential is exactly $V=V_0 (1-\varphi )$, then we have $\varphi_{\text{end}}= 2 a_{p=1} $, but this may be easily modified by higher order terms in $\varphi$.
In this sense, the first order pole inflation is less universal (more model-dependent) than the other cases.
The slow-roll parameters are expressed as $\epsilon = \frac{\varphi_{\text{end}}}{2 a_1}e^{-N/a_1}$ and $\eta=-\frac{1}{2a_1}$, so the inflationary observables are expressed as
\begin{align}
n_{\text{s}}=&1-\frac{1}{a_1}, &  r=& 16 \left(\frac{\varphi_{\text{end}}}{2 a_1} \right) e^{-N/a_1},
\end{align}
at the lowest order of $e^{-N/a_1}$.
Similarly to the case of $0<p<1$, the attractor values of these observables are $n_{\text{s}}\to - \infty$ and $r \to 0$.
As we will see, however, first order pole inflation can be consistent with observation for intermediate values of the attractor parameter $a_1$.

\begin{figure}[htbp!]
\centering
\includegraphics[width=0.75\columnwidth]{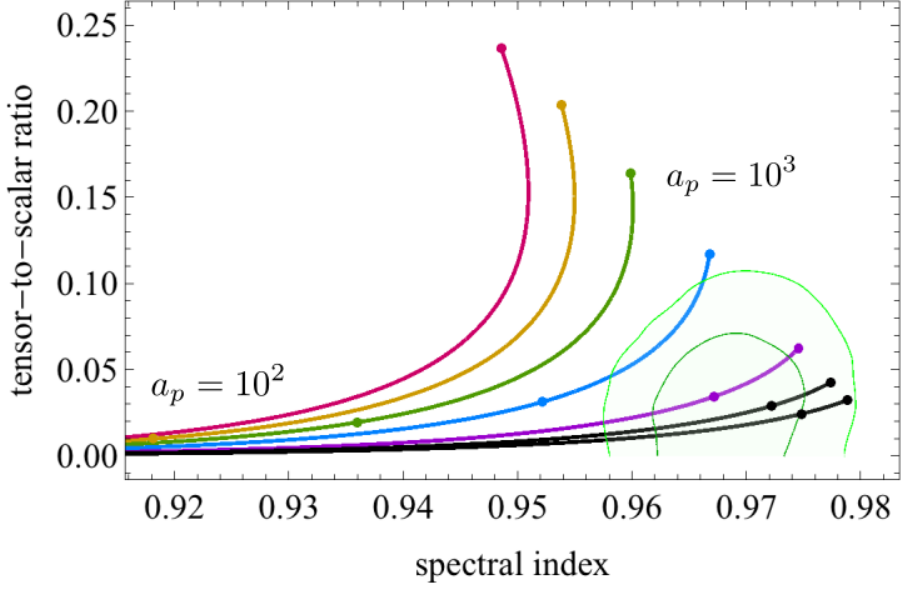}
  \caption{Attractor behavior of first order pole inflation with monomial potentials.
  The power $m$ is taken as $m= 5, \, 4, \,  3, \, 2, \, 1, \, 2/3$, and $1/2$ from top to bottom. 
  The $m=2$ case (blue line) corresponds to natural inflation.
  The dots on the lines correspond to $a_1=10^3$ (upper right) and $10^2$ (lower left).
  The $e$-folding is set to $N=60$.
  The light green lines are the 1 and 2 sigma contours of the Planck TT+lowP+BKP+lensing+BAO+JLA+$H_0$ constraint taken from Fig.~21 of Ref.~\cite{Ade:2015xua}.
  }
 \label{fig:1st}
  \end{figure}

%example leading to natural inflation-type models
As a concrete example, consider the model~\eqref{L_p} with $p=1$.
The canonical inflaton $\phi$ is related to the original one by $\widetilde{\varphi}=\sin (\phi /\sqrt{a_{1} })$.
In terms of the canonical field, eq.~\eqref{L_p} with $p=1$ becomes
\begin{align}
\left( \sqrt{-g}\right)^{-1}\mathcal{L}= - \frac{1}{2} \partial^{\mu} \phi \partial_{\mu} \phi - \lambda_m \sin^m \left( \frac{\phi}{\sqrt{a_{1} }} \right).
\end{align}
Thus, the canonical potential is given by a power of the sinusoidal function.
This is a generalization of the natural inflation potential, $V= V_0 ( 1- \cos (\phi/f)) = 2 V_0 \sin^2 (\phi/2f)$.
The spectral index and tensor-to-scalar ratio are derived as
\begin{align}
n_{\text{s}}= &\frac{\left( -2 m^3 -4 m a_{1}  +m^2 a_{1}  + 2 a_{1}  ^2 \right) - 2a_{1}  \left(m^2+ a_{1}  \right)e^{-2m N /a_{1} }}{a_{1}  \left( \left(m^2 + 2 a_{1}  \right) -2 a_{1} e^{-2m N /a_{1} } \right)}, \\
r = & \frac{16m^2 e^{-2m N /a_{1} }}{\left(m^2+2a_{1}  \right)-2a_{1} e^{-2m N /a_{1} }}.
\end{align}
These are plotted in Fig.~\ref{fig:1st}.
One of the simple cases, $m=1$, has a parameter range well consistent with the observation.
Two cases with fractional power are also shown for illustration, and they are also consistent with the observation.

Note that the attractor parameter $a_1$ can be identified with the square of the decay constant of natural inflation type models.\footnote{
This notion is due to M.~Scalisi though his idea was not in the context of first order pole inflation.
}
In this model, it controls not only the tensor-to-scalar ratio but also significantly controls the value of the spectral index, see Fig.~\ref{fig:1st}.

%%%%%%%%%%%%%%%%%%%%%%%%%%%%%%%%%%%%%%%%%
\section{Effects of other poles in the Lagrangian \label{sec:perturbation}}
For pole inflation with $p\geq 2$, the place of the pole becomes infinitely far from the vacuum in terms of the canonical field.
This makes a vast inflationary plateau, or in other words, an approximate shift symmetry emerges.
For $p<2$, the place of the pole is in a finite distance from the vacuum.
This raises a possibility that the inflaton goes into the region where the sign or phase of the kinetic term of the original field becomes unphysical.
Beyond the pole, we cannot say anything and this point is the boundary of the validity of our effective field theory.
A similar situation was encountered in Ref.~\cite{Ferrara:2013rsa}.
Although inflation occurs within the valid region and eternal inflation can occur, it may be difficult to provide a proper initial condition for inflation without entering such an unphysical region.
We may expect something unusual happens near the boundary of the effective theory, namely near the location of the pole.
We parametrize our ignorance by corrections to the canonical inflaton potential which grows toward the location of the pole.
Such corrections can be expressed in terms of the original field as a pole of order $t$ in the potential or another pole in the kinetic term whose order $q$ is larger than $p$,\footnote{
Up to the first order of $a_q$ and $b_t$, these effects are equivalent just depending on the choice of the field basis~\cite{Nakayama:2016eqv}.
}
\begin{align}
\left( \sqrt{-g}\right)^{-1}\mathcal{L}= -\frac{1}{2} \left( \frac{a_p}{\varphi^p} + \frac{a_q}{\varphi^q} \right)  \partial^{\mu} \varphi \partial_{\mu} \varphi - V_0 \left( \frac{b_t}{\varphi^t} + 1 -\varphi +\mathcal{O}(\varphi^2)  \right).
\end{align}
Similarly to Sec.~\ref{sec:fractional}, the general case with the coefficient $c$ of the linear potential can be recovered by $a_p\to a_p c^{p-2}$, $a_q \to a_q c^{q-2}$, and $b_t \to b_t c^{t}$.
The higher order pole in the kinetic term for $p\geq 2$ can be interpreted as a source of shift symmetry breaking for the canonical inflaton potential~\cite{Broy:2015qna}.
In addition to that, we take into account possible poles in the potential term.

We assume that during last 50 to 60 $e$-foldings in which our observable cosmological scales exit the horizon, the effect of higher order pole is subdominant to the original $p$-th order pole, \textit{i.e.} $  |a_q/\varphi^q| \ll |a_p/\varphi^p|$.  We also require the pole in the potential is subdominant compared to the constant contribution, \textit{i.e.} $|b_t/\varphi^t| \ll 1$.
Thus, we treat $a_q$ and $b_t$ perturbatively.
The corrected spectral index and tensor-to-scalar ratio are
\begin{align}
n_{\text{s}}=& n_{\text{s}, \, 0}+ \delta_{\text{kin}}n_{\text{s}} + \delta_{\text{pot}}n_{\text{s}}, & 
r = & r_0 +\delta_{\text{kin}}r + \delta_{\text{pot}}r,
\end{align}
where $n_{\text{s}, \, 0}$ and $r_0$ are given by eq.~\eqref{nsr_p}.
Up to the first order, the correction from the $q$-th order pole in the kinetic term is~\cite{Broy:2015qna}
\begin{align}
\delta_{\text{kin}}n_{\text{s}}=& - \frac{(q-p)(q-p-1)a_q}{(q-1)a_p^2} \left( \frac{p-1}{a_p} N \right)^{\frac{q-2p+1}{p-1}} , \label{ns_q} \\
\delta_{\text{kin}}r = & - \frac{8(q-p-1)a_q}{(q-1)a_p^2}\left( \frac{p-1}{a_p} N \right)^{\frac{q-2p}{p-1}} . \label{r_q}
\end{align}
The correction from the $t$-th order pole in the potential is
\begin{align}
\delta_{\text{pot}} n_{\text{s}}=& \frac{t(t+1)(p+2t)b_t}{(p+t) a_p} \left( \frac{p-1}{a_p} N\right)^{\frac{t-p+2}{p-1}} , \label{ns_t} \\
\delta_{\text{pot}} r =& \frac{8t (p+2t) b_t}{(p+t)a_p} \left( \frac{p-1}{a_p}N \right)^{\frac{t-p+1}{p-1}}. \label{r_t}
\end{align}
Note that even if there is a pole in the potential ($t\geq 1$), $n_{\text{s}}$ and $r$ do not receive corrections of positive power of the $e$-folding number $N$ if $t\leq p-2$.
This happens only for $p\geq 3$.
Thus, if the order of the pole in the kinetic term is sufficiently higher than the order of the pole in the potential, the effect of the latter on the observables is small.

When $p=2$, corresponding to $\alpha$-attractor, eqs.~\eqref{ns_q} and \eqref{r_q} become~\cite{Broy:2015qna}
\begin{align}
\delta_{\text{kin}}n_{\text{s}}=& - \frac{(q-2)(q-3)a_q}{(q-1)a_{p=2}^{q-1}}N^{q-3}, \\
\delta_{\text{kin}}r = & -\frac{8(q-3)a_q}{(q-1)a_{p=2}^{q-2}}N^{q-4}.
\end{align}
Similarly, eqs.~\eqref{ns_t} and \eqref{r_t} for $p=2$ become
\begin{align}
\delta_{\text{pot}} n_{\text{s}}=& \frac{2 t (t+1)^2 b_t}{(t+2) a_{p=2}^{t+1}}  N^{t} ,\\
\delta_{\text{pot}} r =& \frac{16t (t+1) b_t}{(t+2)a_{p=2}^t} N^{t-1}.
\end{align}

In the case of first order pole inflation ($p=1$), the counterparts are
\begin{align}
\delta_{\text{kin}} n_{\text{s}} =& -\frac{(q-2)a_q}{a_{p=1}^2 }\left( \frac{ e^{N/a_{p=1}} }{\varphi_{\text{end}}} \right)^{q-1}, \\
\delta_{\text{kin}} r = & - \frac{8 (q-2)a_q}{(q-1)a_{p=1}^2 } \left( \frac{e^{N/a_{p=1}} }{\varphi_{\text{end}}} \right)^{q-2},
\end{align}
and
\begin{align}
\delta_{\text{pot}} n_{\text{s}} =& \frac{t(2t+1) b_t}{a_{p=1}} \left( \frac{e^{N/a_{p=1}}}{\varphi_{\text{end}}} \right)^{t+1}, \\
\delta_{\text{pot}} r=& \frac{8t (2t+1) b_t}{(t+1)a_{p=1}}\left( \frac{e^{N/a_{p=1}}}{\varphi_{\text{end}}}\right)^t. 
\end{align}
One can see that the first order pole inflation is sensitive to the corrections.  Namely, these corrections depend exponentially on the $e$-folding number.

%%%%%%%%%%%%%%%%%%%%%%%%%%%%%%%%%%%%%%%%%
\section{Pole inflation with a singular potential \label{sec:singular}}
In the above, we have seen that a pole in the kinetic term whose order is high enough makes the effects of a pole in the potential small.
We extend this further and consider a potential which mainly consists of a singular part.\footnote{
If the Jordan frame function $\Omega_{\text{J}}$ is responsible for the pole of the kinetic term, it has also a pole in the potential, see eq.~\eqref{frames_relation}.
Even if $\Omega_{\text{J}}^{-1}$ does not have a pole, presence of poles both in the kinetic and potential terms is naturally obtained in supergravity because the poles in the kinetic term are originated from K\"{a}hler potential, and it also controls both $F$-term and $D$-term potentials. (The author thanks K.~Nakayama for pointing this out.) Without tuning of superpotential, poles in the potential are generically expected when the kinetic term has poles.
For example, a simple $D$-term model (3.1) in Ref.~\cite{Nakayama:2016eqv} has second order poles both in $\Omega'{}^2_{\text{J}}$ and $V_{\text{J}}$ (and hence in $K_{\text{E}}$ and $V_{\text{E}}$).
Also, removing the ad hoc factor $(3-\Phi^2)^{(3\alpha-1)/2}$ from the superpotential of the superconformal $\alpha$-attractor ($F$-term model) leads to a $(3\alpha -1)$-th order pole in the potential, see eq.~(4.3) of Ref.~\cite{Kallosh:2013yoa}.
}
That is, $b_t/\varphi^t$ term is no longer perturbation, but it is the main part of the potential,
\begin{align}
\left( \sqrt{-g}\right)^{-1}\mathcal{L}= -  \frac{a_p}{2 \varphi^p}  \partial^{\mu} \varphi \partial_{\mu} \varphi - \frac{C}{\varphi^s}\left( 1 + \mathcal{O}(\varphi)  \right), \label{Lsingular}
\end{align}
where $C$ is an overall constant of the potential, and $s(>0)$ denotes the order of the strongest pole in the potential relevant during inflation for the observable scales.

\begin{figure}[htb]
 \centering
  \subcaptionbox{original potential.\label{sfig:dv_pt_org}}
{\includegraphics[width=0.32\columnwidth]{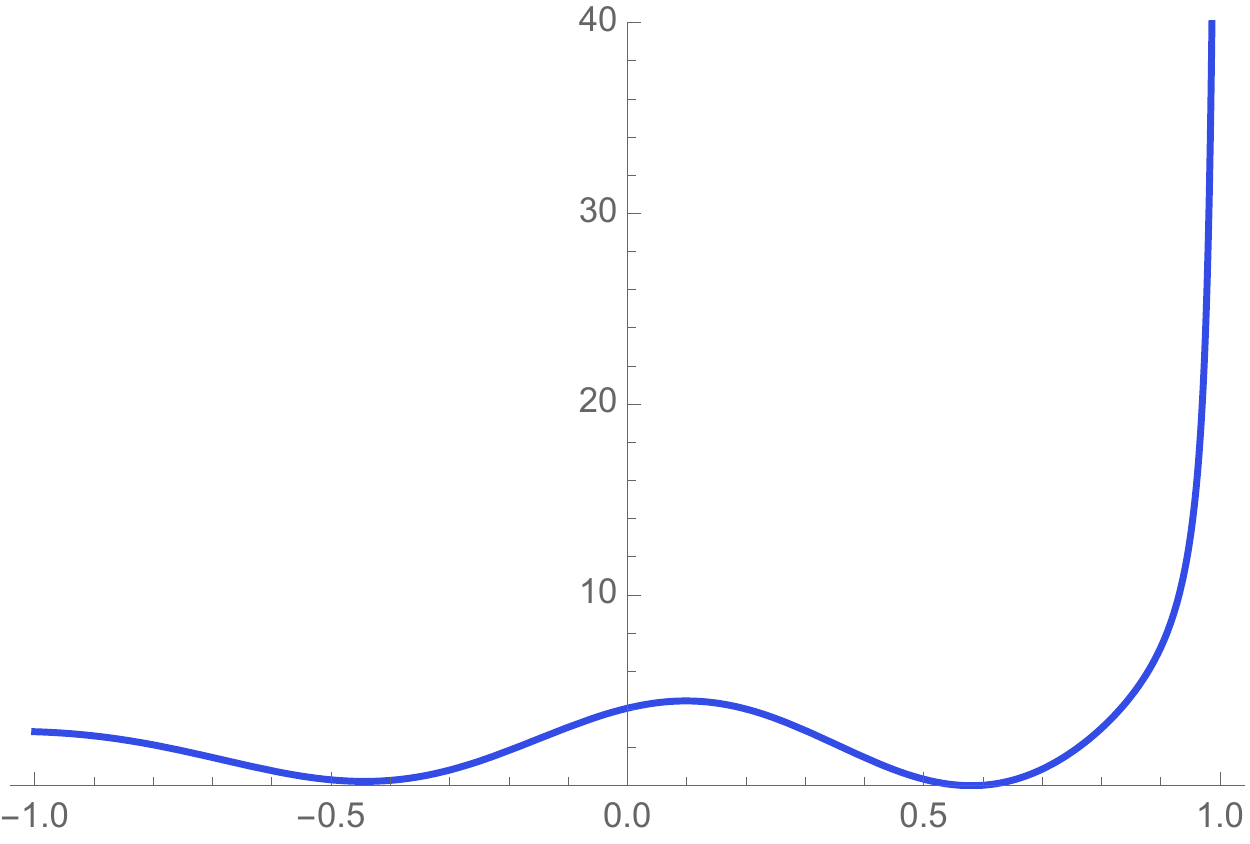}}~
 \subcaptionbox{$p=2$: power-law.\label{sfig:dv_pt_powerlaw}}
{\includegraphics[width=0.32\columnwidth]{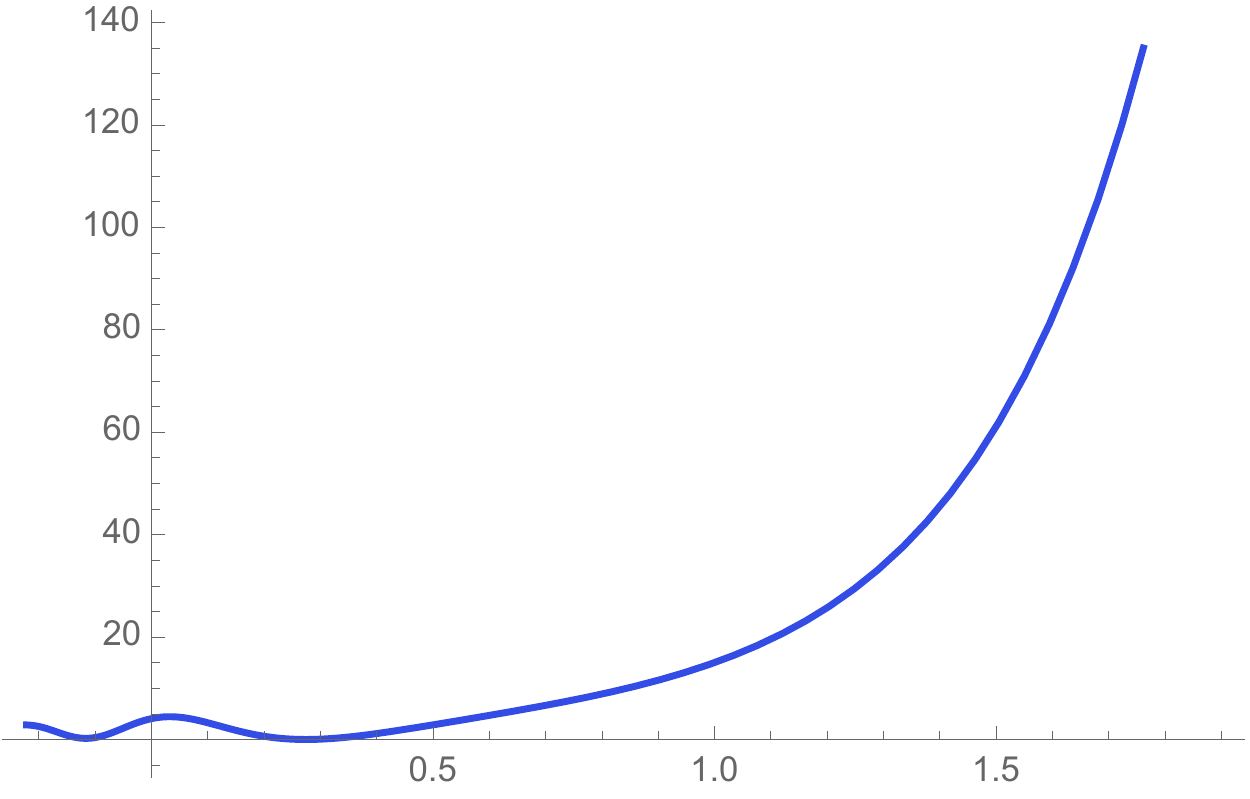}}~
 \subcaptionbox{$p>2$: monomial chaotic. \label{sfig:dv_pt_chaotic}}
{\includegraphics[width=0.32\columnwidth]{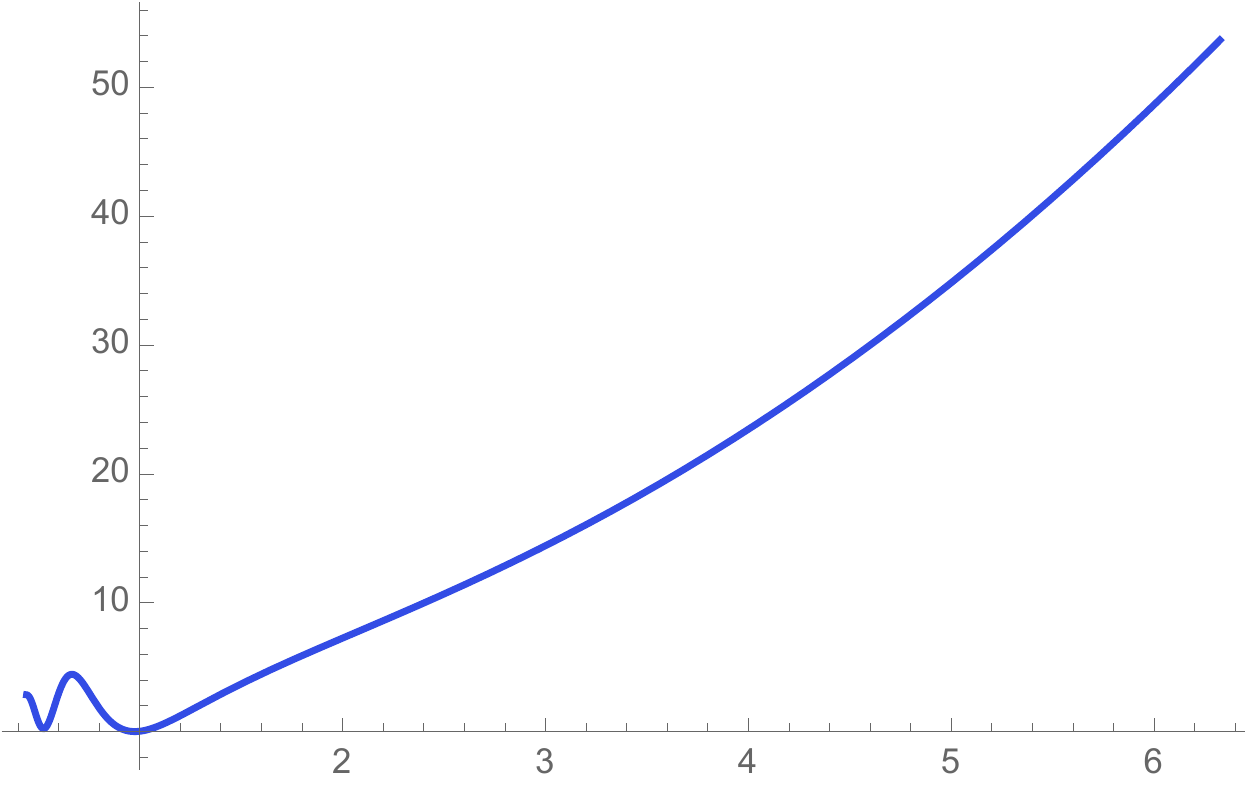}}~

  \caption{Effects of canonical normalization for a divergent potential.
  (\subref{sfig:dv_pt_org}) The original divergent potential before canonical normalization.
  (\subref{sfig:dv_pt_powerlaw}) The canonical potential for $p=2$, which asymptotes to an exponential function leading to power-law inflation.
  (\subref{sfig:dv_pt_chaotic}) The canonical potential for $p>2$, which asymptotes to a monomial function.}
 \label{fig:dv_pt}
  \end{figure}

The canonical potential is 
 \begin{align}
 V= \begin{cases}
 C \left( \frac{p-2}{2\sqrt{a_p}}\phi \right)^{\frac{2s}{p-2}} + \cdots  & (p \neq 2), \label{Vcan_s} \\
 C e^{s \phi/\sqrt{a_p}} + \cdots & (p=2).
 \end{cases} 
 \end{align}
Thus, for $p\neq 2$, we obtain an effectively monomial potential for chaotic inflation.
For example, if we take $s=1$, then $p=3, \, 4$, and 5 lead to canonical potentials with power $2, \, 1$, and $2/3$, respectively.  Higher order poles in the kinetic term result in smaller fractional power of the canonical potential.
Note also that the presence of a pole in the potential in the case of $\alpha$-attractor ($p=2$) results in an exponential function whose exponent depends on the order of the pole $s$ as well as the attractor parameter $a_{p=2}=3\alpha /2$.
It leads to power-law inflation~\cite{Abbott:1984fp, Lucchin:1984yf}, but it has been excluded.
Also, we do not consider the case $p<2$ in this section since it does not lead to a suitable inflaton potential.
These effects of canonical normalization are visually presented in Fig.~\ref{fig:dv_pt}.
The relation between the field and the $e$-folding is now
\begin{align}
N = \frac{a_p}{s(p-2)}\left( \frac{1}{\varphi_N^{p-2}} - \frac{1}{\varphi_{\text{end}}^{p-2}} \right).
\end{align}
The spectral index and tensor-to-scalar ratio in the attractor limit ($a_p \to 0$) are
\begin{align}
n_{\text{s}}=&  1- \frac{p+s-2}{(p-2)N}, &   r=& \frac{8s}{(p-2)N}. \label{nsr_chaotic}
\end{align}
These results are consistent with Ref.~\cite{Rinaldi:2015yoa}, and taking $s=(p-2)/2$ reproduces one of their main results, $r= 8(1-n_{\text{s}})/3$.

If we increase $a_p$, the relative importance between terms of different power changes, and terms represented by dots in eq.~\eqref{Vcan_s} become important.  In such a case, the inflaton potential becomes a polynomial.
Choosing parameter values properly, polynomial chaotic inflation can fit the observational data well~\cite{Destri:2007pv, Nakayama:2013jka, *Nakayama:2013txa}.

\begin{figure}[htbp!]
\centering
\includegraphics[width=0.75\columnwidth]{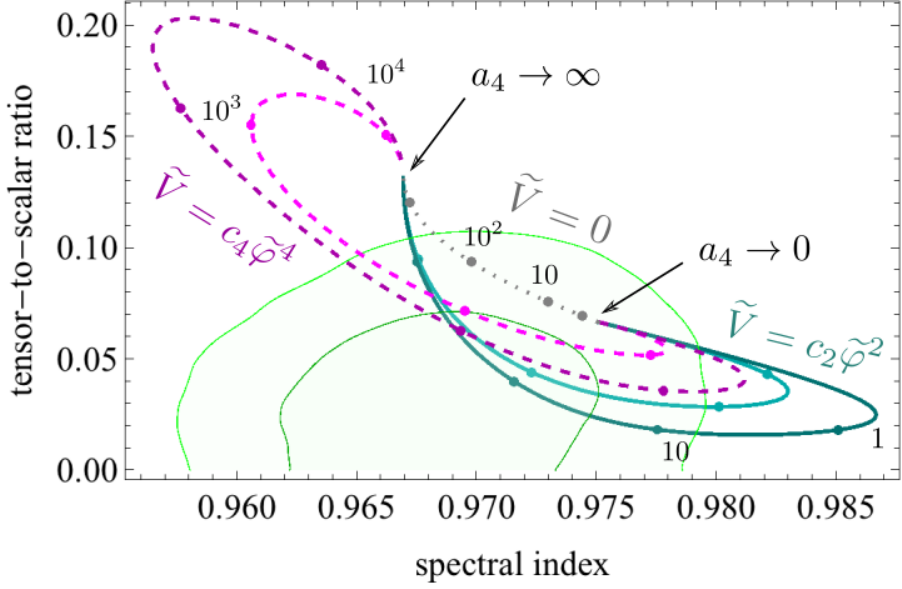}
  \caption{Attractor behavior of pole inflation with singular potentials.
  See eq.~\eqref{CA_example}. 
  The orders of poles are taken as $p=4$ and $s=1$.
The potentials for the gray dotted, cyan solid, and magenta dashed lines are taken as $\widetilde{V}= 0, \, c_2 \widetilde{\varphi}^2$, and $c_4 \widetilde{\varphi}^4$, respectively, with $c_2=30$ (light color, short curve) and 100 (deep color, long curve) and $c_4=10$ (short) and 30 (long).  
In the small $a_4$ limit, the predictions converge at the point of linear potential, and at the large $a_4$ limit, they are attracted to the point of quadratic potential (the double attractor).
 The dots on the gray and cyan lines correspond to $a_4=1, \, 10, \, 10^2$, and $10^3$, whereas the dots on the magenta lines correspond to $a_4=10, \, 10^2, \, 10^3$, and $10^4$.
  The $e$-folding is set to $N=60$.
  The Planck contours are same as those in Fig.~\ref{fig:1st}.
  }
 \label{fig:singular}
  \end{figure}

We emphasize that the Lagrangian~\eqref{Lsingular} leading to eq.~\eqref{nsr_chaotic} is also an attractor.
To see this explicitly, consider the following example,
\begin{align}
\left( \sqrt{-g}\right)^{-1}\mathcal{L}= -\frac{1}{2} \frac{a_4}{(1-\widetilde{\varphi}^2)^4}  \partial^{\mu} \widetilde{\varphi} \partial_{\mu} \widetilde{\varphi} - C\left(\frac{2 \widetilde{\varphi}^2}{1-\widetilde{\varphi}^2} + \widetilde{V}(\widetilde{\varphi}) \right), \label{CA_example}
\end{align}
where $C$ is an overall coefficient, and $\widetilde{V}(\widetilde{\varphi})$ is a some function regular in $-1\leq \widetilde{\varphi} \leq 1$.
In this model, there are fourth order poles ($p=4$) at $\widetilde{\varphi}=\pm 1$ in the kinetic term and first order pole ($s=1$) at the same positions in the potential, so the linear potential for the canonical field is obtained at the attractor limit.
The attractor behavior of this model with $\widetilde{V}(\widetilde{\varphi} )=0, \, c_2 \widetilde{\varphi}^2$, and $c_4 \widetilde{\varphi}^4$ is shown in Fig.~\ref{fig:singular}.
It is seen that all lines converge at the attractor point \eqref{nsr_chaotic} for $a_4 \to 0$.
At the opposite limit, $a_4 \gg 1$, all the lines are attracted to the prediction of the quadratic potential.
This is a realization of the double attractor mechanism~\cite{Kallosh:2014rga, Kallosh:2014laa, Mosk:2014cba}.
It should be stressed that the chaotic inflation limit in eq.~\eqref{nsr_chaotic} has nothing to do with chaotic inflation limit in Ref.~\cite{Mosk:2014cba}, which is instead related to the opposite (quadratic model) limit in Fig.~\ref{fig:singular}.

Note that there is no global (approximate) shift symmetry in terms of the canonical field $\phi$ in eq.~\eqref{Vcan_s} which would protect flatness of the canonical potential.
It implies that there is no corresponding symmetry for $\varphi$.
Then, it is likely that there are additional poles in the Lagrangian which have non-negligible effects.
We can calculate corrections to eq.~\eqref{nsr_chaotic} whose origin is a higher order pole either in the kinetic term or in the potential.
Consider the following Lagrangian,
\begin{align}
\left( \sqrt{-g}\right)^{-1}\mathcal{L}= -\frac{1}{2} \left( \frac{a_p}{\varphi^p} + \frac{a_q}{\varphi^q} \right)  \partial^{\mu} \varphi \partial_{\mu} \varphi - \frac{C}{\varphi^s} \left( \frac{b_t}{\varphi^t} + 1 + \mathcal{O}(\varphi)  \right).
\end{align}
Similarly to the previous case, we consider the situation in which $|a_q/\varphi^q| \ll |a_p / \varphi^p|$ and $|b_t/\varphi^t|\ll 1$ are satisfied, and these small quantities are treated perturbatively.
The corrections are,
\begin{align}
\delta_{\text{kin}} n_{\text{s}} = & - \frac{s(q-p)(q-p-s)a_q}{(q-2)a_p^2} \left( \frac{s(p-2)}{a_p} N \right)^{\frac{q-2p+2}{p-2}} , \\
\delta_{\text{kin}} r=& -\frac{8 s^2 (q-p) a_q}{(q-2)a_p^2} \left( \frac{s(p-2)}{a_p} N \right)^{\frac{q-2p+2}{p-2}},
\end{align}
and
\begin{align}
\delta_{\text{pot}} n_{\text{s}} = & \frac{t(t-s)(p+2t-2)b_t}{(p+t-2)a_p} \left( \frac{s (p-2)}{a_p} N \right)^{\frac{t-p+2}{p-2}}   ,\\
\delta_{\text{pot}} r=&  \frac{8st(p+2t-2)b_t}{(p+t-2)a_p} \left( \frac{s(p-2)}{a_p} N \right)^{\frac{t-p+2}{p-2}}  .
\end{align}
In general, these corrections may not be small since there is no symmetry for $\varphi$ as mentioned above.
In this sense, the inflationary attractors with a monomial potential for chaotic inflation is less universal than the plateau type attractors such as $\alpha$-attractor.
This is similar for the case of $p<2$ with or without a pole in the potential.
For such cases, locally flat canonical potentials suitable for slow-roll inflation and their underlying pole structures in the original field $\varphi$ are viewed as accidental ones, which may be justified by anthropic arguments.

%%%%%%%%%%%%%%%%%%%%%%%%%%%%%%%%%%%%%%%%%
\section{Conclusion \label{sec:summary}}
Inflationary attractors or pole inflation is an interesting mechanism which universally leads to inflationary observables consistent with the cosmological data.
We explicitly demonstrate the attractor behavior of pole inflation with various pole orders $p$ taking a monomial potential in the original variable as a simple example.
We find that the first order pole inflation can lead to variants of natural inflation.  This may depend on the global structure of poles in the original kinetic term.
The decay constant of the natural inflation model can be identified as the square root of the attractor parameter, which is the residue of the pole.
We discussed the issue of the initial conditions and validity of the effective theory, and considered the effects of terms growing toward the boundary of the theory, namely additional poles.
These poles may be either in the kinetic term or in the potential.
The corrections from these terms to the inflationary observables $(n_{\text{s}}, \, r)$ have been calculated.
Moreover, we explored the possibility that inflation happens on a singular potential.
This leads to the inflationary attractors whose canonical potential is a monomial potential.
Thus, the notion of inflationary attractors and pole inflation are generalized to include the \textit{sinusoidal attractor}, the \textit{power-law attractor}, and the \textit{chaotic attractors} in addition to the \textit{hilltop} or \textit{plateau attractors}.
Note that most of universality classes of inflation~\cite{Mukhanov:2013tua, Roest:2013fha, Garcia-Bellido:2014gna, Binetruy:2014zya} can be realized in the context of pole inflation, see Table~\ref{tab:univ}.
Pole inflation in general can thus be viewed as a concrete realization of the universality classes of inflation.
This will deepen our understanding of inflationary models and mechanisms of inflationary attractors.
It will be interesting to explore formulation in theories with non-minimal coupling to gravity and possible connections to ultraviolet theories.  Some basic analyses in these lines can be found in Refs.~\cite{Broy:2015qna, Rinaldi:2015yoa}.

\begin{table}[htb]
{\footnotesize 
\begin{center}
\caption{Correspondence with universality classes of inflation.  This is an extension of Table 1 in Ref.~\cite{Garcia-Bellido:2014gna}.
The $\epsilon(N)$ is the dependence of the slow-roll parameter $\epsilon$ on the $e$-folding number $N$, and $V(\phi)$ is a canonical potential.  All the coefficients are omitted emphasizing the rough functional structures.
The power $n$ in the perturbative $k=1$ class (chaotic model) is related to the orders of the poles $p$ in the kinetic term and $s$ in the potential as $n=2s/(p-2)$ (\textit{cf}.~eq.~\eqref{Vcan_s}).
The power $k$ of the perturbative classes is related to the power $n$ in the canonical potential as $k= \frac{2(|n|-1)}{|n|-2}$, which is further related to the pole order $p$ as $n= - \frac{2}{p-2}$ (\textit{cf}.~eq.~\eqref{Vcan}). 
}
  \begin{tabular}{|c|c|c|c|c|}
\hline
universality classes & $\epsilon(N)$ & $V(\phi)$ & inflation model & corresponding pole \\
\hline
\hline
Constant & constant & $e^{\phi}$ & Power-law & $p=2$ w/ a singular potential \\ \hline
Perturbative $k=1$ &  $1/N$ & $\phi^n \quad (n>0)$ & Chaotic & $p>2$ w/ a singular potential \\ \hline
Perturbative $1<k<2$ &  $1/N^k$ & $ 1 - \phi^n \quad (n<0)$ & Inverse-Hilltop & $p>2$ \\ \hline
Perturbative $k=2$ &  $1/N^2$ & $1-e^{-\phi}$ & Starobinsky & $p=2$ \\ \hline
Perturbative $k>2$ &  $1/N^k$ & $1-\phi^n \quad (n>0)$ & Hilltop & $1<p<2$ \\ \hline
Non-Perturbative &  $e^{-N}$ & $1-\phi^2 $ & (Natural) & $p=1$ \\ \hline
Logarithmic & $(\ln N )/ N$ & $1 - \phi e^{-\phi}$ & K\"ahler Moduli & $p=2$ w/ log.~corrections \\
\hline
  \end{tabular}
  \label{tab:univ}
  \end{center}
  }
\end{table}

%%%%%%%%%%%%%%%%%%%%%%%%%%%%%%%%%%%%%%%%%
\section*{Acknowledgments}
The author is grateful to T.~Asaka, S.~Iso, H.~Kawai, K.~Kohri, and T.~Noumi for discussion on classification of flat potentials, K.~Nakayama, K.~Saikawa, and M.~Yamaguchi for discussion on pole inflation in supergravity, T.~Kitahara and N.~Kitajima for useful comments on the manuscript, and M.~Scalisi for useful comments on the manuscript and discussion on the relation between attractors and natural inflation. 
 He was supported by the Grant-in-Aid for JSPS Fellows and the Grant-in-Aid for Scientific Research  No.~26$\cdot$10619.

\bibliographystyle{utphys}
\bibliography{ref2.bib}
\end{document}